\documentclass[12pt]{article}
\usepackage[dvips]{graphicx,color}
\usepackage{amsfonts}
\usepackage{amsmath}
\begin{document}
\title{On the induced gauge invariant mass}
\author{C.~D.~Fosco$^a$~\footnote{fosco@cab.cnea.gov.ar},
  L.~E.~Oxman$^b$~\footnote{oxman@dft.if.uerj.br},
  S.~P.~Sorella$^b$~\footnote{sorella@uerj.br}
  \\
  {\normalsize\it $^a$Centro At\' omico Bariloche - Instituto Balseiro,}\\
  {\normalsize\it Comisi{\'o}n Nacional de Energ\'{\i}a At{\'o}mica}\\
  {\normalsize\it 8400 Bariloche, Argentina.}
  \\
  {\normalsize\it $^b$UERJ - Universidade do Estado do Rio de
    Janeiro,}\\
  {\normalsize\it Rua S\~{a}o Francisco Xavier 524, 20550-013,}\\
  {\normalsize\it Maracan\~{a}, Rio de Janeiro, Brazil.}}

\date{\today}
\maketitle
\begin{abstract}
\noindent
We derive a general expression for the gauge invariant mass ($m_G$)
for an Abelian gauge field, as induced by vacuum polarization, in
$1+1$ dimensions.  From its relation to the chiral anomaly, we show
that $m_G$ has to satisfy a certain quantization condition. This
quantization can be, on the other hand, explicitly verified by using
the exact general expression for the gauge invariant mass in terms of
the fermion propagator.  This result is applied to some explicit
examples, exploring the possibility of having interesting physical
situations where the value of $m_G$ departs from its canonical value.
We also study the possibility of generalizing the results to the $2+1$
dimensional case at finite temperature, showing that there are indeed
situations where a finite and non-vanishing gauge invariant mass is
induced.
\end{abstract}

\section{Introduction}

Some important physical quantities displaying quantization properties,
may sometimes be represented by means of {\em momentum\/} space integrals
which exhibit their topologically invariant character. Considerable
effort has been devoted to find these representations, since they are
often very useful to prove their quantization, as well as their
stability under perturbations.  

This is the case, for instance, of the transverse conductivity $\sigma$ in
$QED_3$, which can be represented as a momentum space integral given by
\cite{mats,mats1}
\begin{equation}\label{eq:ihall}
\sigma = \frac{1}{3!}i e^2 \varepsilon^{\mu\nu\rho} \int \frac{d^3p}{(2\pi)^3} {\rm Tr} \left[ 
\partial_{\mu}S^{-1} S  \partial_{\nu}S^{-1} S    \partial_{\rho}S^{-1} S \right] \ ,
\end{equation}
where $S(p)$ is the full fermion propagator. As shown in \cite{mats},
due to the ultraviolet behavior of the fermion propagator, this
integral reduces to the Kronecker topological invariant which labels
the homotopy classes in $\Pi_2(S^2)$, that is, $\sigma$ is quantized.  A
similar representation for the induced Chern-Simons coefficient has
been obtained in~\cite{cl} by using a cubic lattice regularization for
the Euclidean fermionic action. In this case, the possible values for
the Chern-Simons coefficient turn out to be labeled by the winding
number characterizing the mapping between the three-dimensional torus
in momentum space and the normalized quaternion corresponding to the
fermion propagator: $S(p)/\sqrt{det(S)}$.

Topology in momentum space has also been advocated in~\cite{volovik},
in order to discuss the stability of neutrino masses in the Standard
Model and the spectrum of excitations in effective two dimensional
models such as Helium-3. On the other hand, in the context of
Yang-Mills theories, one of the main open problems is that of
understanding the non-perturbative generation of a mass gap for the
gauge fields as a consequence of the dynamics in the infrared regime.
This problem is of crucial importance for confinement in QCD, and for
analyzing finite temperature effects. For instance, in~\cite{rj}, a
gap equation, based on the introduction of gauge invariant mass terms,
has been proposed and applied to the Yang-Mills theory in $2+1$
dimensions.

The aim of this work is to obtain a useful momentum space
representation for detecting the existence of an induced gauge
invariant mass, $m_G$, for an Abelian gauge field. In particular, we
shall be able to derive a general expression for $m_G$, valid for
different space-time dimensions and geometries in momentum space.

The paper is organized as follows. Section 1 is devoted to a short
review of the well-known relationship between the induced gauge
invariant mass and the chiral anomaly in $1+1$ dimensions. In section
2, a general momentum space representation for the induced gauge
invariant mass is provided.  Section 3 is devoted to the applications
of the aforementioned representation to the case of square lattice
geometry and to the case of finite temperature in $2+1$ dimensions.
Section~\ref{sec:conc} presents our conclusions.

\section{Relationship Between the Gauge Invariant Mass and the Chiral
  Anomaly in $1+1$ dimensions} 

By gauge invariant mass, $m_G$, we understand the value of the
constant appearing in a gauge invariant mass term for an Abelian gauge
field.  In $d+1$ dimensional Euclidean spacetime, a gauge invariant
mass term Lagrangian ${\mathcal L}_m$ is given explicitly by
\begin{equation}\label{eq:defmt}
{\mathcal L}_m \;=\; \frac{1}{2} m^2_G \, A_\mu \delta^\perp_{\mu\nu} A_\nu 
\end{equation}
where $\delta^\perp_{\mu\nu}=\delta_{\mu\nu}-\frac{\partial_\mu\partial_\nu}{\partial^2}$ is the transverse Kronecker
$\delta$ in $D=d+1$ dimensions.

This mass term for $A_\mu$ is explicitly gauge invariant, although, of
course, at the price of introducing a non-locality in the action.
That is often the reason for not including this term in the classical
Lagrangian of a standard local quantum field theory. In spite of
this fact, terms like this naturally arise when evaluating radiative
corrections to the effective action.  In {\em massless\/} $QED$ in
$1+1$ dimensions, i.e., the Schwinger model~\cite{schwinger}, it is
induced by the one-loop vacuum polarization graph.  Moreover, for such
a model, the one-loop result is exact, since it does not suffer from
higher order corrections. These results can also be shown to be
related to the chiral anomaly in $1+1$ dimensions, whence the
non-renormalization of $m_G$ is inherited. For this model, the
non-locality can be made good since, in the bosonization approach, the
gauge field may be written in terms of two scalar fields: $A_\mu =
\epsilon_{\mu\nu} \partial_\nu \sigma + \partial_\mu\varphi$. In terms of these scalars, the gauge invariant
mass term becomes a standard, local, mass term for the $\sigma$ field.

In the $1+1$ dimensional case, the value of $m_G$ for a single
fermionic flavor has the well-known value $m^2_G={e^2}/{\pi}$.  In
this article, we are interested in deriving general expressions for
$m_G$, as a first step to extending the useful $1+1$ dimensional
result into several non trivial directions.  In order to define the
problem more precisely, we write $m_G$ in terms of the exact vacuum
polarization tensor~\footnote{The tilde denotes the momentum space
  version of an object, whenever it its convenient to distinguish it
  from its coordinate space version.} ${\tilde \Pi}_{\mu\nu}(k)$. The Ward
identity for ${\tilde \Pi}_{\mu\nu}(k)$, $k_\mu{\tilde \Pi}_{\mu\nu}(k)=0$, allows
us to write:
\begin{equation}\label{eq:pimn}
{\tilde \Pi}_{\mu\nu}(k)\;=\; {\tilde \Pi}(k^2) \; \delta^\perp_{\mu\nu}(k) 
\end{equation} 
where ${\tilde \Pi}(k^2)$ is a scalar function. Thus $m_G$ may also be
obtained as:
\begin{equation}\label{eq:pi0}
m^2_G \;=\; \lim_{k \to 0} \, {\tilde \Pi}(k^2) \;.
\end{equation}
This definition implicitly assumes gauge and Lorentz invariance, two
conditions that, except explicit indication on the contrary, shall be
maintained in everything that follows.

We shall be first concerned with models that can be described by an
Euclidean action with the following structure:
\begin{equation}\label{eq:defs}
S\;=\; \int d^Dx \, {\mathcal L} \;\;\;\;\;\; {\mathcal L}\,=\, {\mathcal L}_F + {\mathcal L}_G
\end{equation}
where ${\mathcal L}_F$ and ${\mathcal L}_G$ denote the fermion and
gauge field Lagrangians, respectively.

Let us begin with a consideration of the $1+1$ dimensional case, in
the simplest non-trivial situation of massless $QED(1+1)$. This will
amount to a re-derivation of known results, although we shall present
them here from a different perspective. The gauge field Lagrangian is
\begin{equation}\label{eq:deflg}
{\mathcal L}_G \,=\, \frac{1}{4} F_{\mu\nu} F_{\mu\nu} \;\;\;,\;\;\; 
F_{\mu\nu} = \partial_\mu A_\nu - \partial_\nu A_\mu \;,
\end{equation}
while fermionic matter is described by a massless Dirac field $\psi$,
with a Lagrangian:
\begin{equation}\label{eq:deflf}
{\mathcal L}_F \;=\; {\bar\psi} ( \not \! \partial + i e \not \!\! A) \psi \;,
\end{equation}
where we adopted the conventions:
$$
(\gamma_\mu)^\dagger \,=\, \gamma_\mu \;\;\; \gamma_5^\dagger \,=\, \gamma_5 \;\;,\;\; \gamma_\mu \gamma_\nu \,=\,
g_{\mu\nu} \,+\, i \, \epsilon_{\mu\nu} \gamma_5
$$
\begin{equation}
g_{\mu\nu}\,=\, \delta_{\mu\nu} \;\;\;\; \epsilon_{01}\,=\,+1 \;.
\end{equation}

The exact vacuum polarization tensor ${\tilde \Pi}_{\mu\nu}$ is, as usual,
defined in terms of the connected gauge field two-point function
$G_{\mu\nu}$ by:
\begin{equation}\label{eq:defpi}
\langle A_\mu A_\nu\rangle_{conn} \,=\, G_{\mu\nu} \;\;\;,
\;\;\; G^{-1}_{\mu\nu} (k) \,=\, k^2 \delta^\perp_{\mu\nu}
\,+\, {\tilde \Pi}_{\mu\nu}(k) \;.  
\end{equation}
It is easy to see that there is, indeed, a relation between $m_G$ and
the chiral anomaly ${\mathcal G}(A)$, since the former may actually be
derived from the latter, at least for massless $QED$ in $1+1$
dimensions. To that end, we define ${\mathcal G}(A)$ by
\begin{equation}\label{eq:defanom}
\partial_\mu \langle J_\mu^5 \rangle \;=\; {\mathcal G}({\mathcal A})\;\;\;;
\end{equation}
where $\langle J_\mu^5 \rangle$ denotes the quantum average of the axial
current \mbox{$J_\mu^5\equiv -i e {\bar\psi}\gamma_5 \gamma_\mu \psi$} in the presence of an external
gauge field ${\mathcal A_\mu}$ (we use the cursive ${\mathcal A}_\mu$
notation to distinguish it from the {\em dynamical\/} gauge field
$A_\mu$).  Hence, the integrated form of the anomaly \cite{anom} is
\begin{equation}\label{eq:defanom1}
\int d^2x \, {\mathcal G}({\mathcal A}) \;=\; \int d^2x \, \partial_\mu\langle J_\mu^5 \rangle \;=\;  
\int d^2x \, \epsilon_{\mu\nu} \partial_\mu \langle J_\nu \rangle 
\end{equation}
where we used the relation: $J_\mu^5 = \epsilon_{\mu\nu} J_\nu$, and $J_\mu\equiv e {\bar\psi}
\gamma_\mu \psi$ is the vector current. Then, from our knowledge 
of the chiral anomaly, we may use the fact that ${\mathcal G}({\mathcal A})$ 
is linear in ${\mathcal A}$ to adopt the linear approximation for 
$\langle J_\nu \rangle$ in (\ref{eq:defanom1}), namely 
\begin{equation}\label{eq:linear}
\langle J_\mu(x) \rangle \;=\; 
\int d^2y \; \Pi_{\mu\nu}(x-y) {\mathcal A}_\nu(y)  
\end{equation}
with $\Pi_{\mu\nu}(x-y)$ denoting the coordinate space vacuum polarization
function.  It should be noted that the $\Pi_{\mu\nu}(x-y)$ appearing in
(\ref{eq:linear}) is exact, since only an expansion in the external field
has been performed.

Inserting then (\ref{eq:linear}) into (\ref{eq:defanom}) and Fourier
transforming, we find that:
\begin{equation}\label{eq:anom1}
\int d^2x \; {\mathcal G}({\mathcal A}) \;=\; \lim_{k\to0} {\tilde \Pi}(k^2) \;\;
\lim_{k\to0} \epsilon_{\mu\nu} i k_\mu {\tilde {\mathcal A}}_\nu (k) \;, 
\end{equation}
assuming that both limits exist. Taking (\ref{eq:pi0}) into account, 
this amounts to:
\begin{equation}\label{eq:anom2}
\int d^2x \; {\mathcal G}({\mathcal A}) \;=\; m^2_G \, 
\Phi({\mathcal A}) \;, 
\end{equation}
where $\Phi({\mathcal A})= \int d^2x \epsilon_{\mu\nu} \partial_\mu{\mathcal A}_\nu$ is the total
flux of $\epsilon_{\mu\nu} \partial_\mu{\mathcal A}_\nu$ in Euclidean spacetime. Equation
(\ref{eq:anom2}) yields $m_G$ in terms of the anomaly. Of course, the
left hand side in equation (\ref{eq:anom2}) being the spacetime
integral of the anomaly,  will have the form of a coefficient
times $\Phi({\mathcal A})$, namely:
\begin{equation}
\int d^2x \, {\mathcal G}({\mathcal A}) \;=\; \xi \, \Phi({\mathcal A})
\end{equation}
thus the content of (\ref{eq:anom2}) is that the coefficient
$\xi={e^2}/{2\pi}$ gives precisely the value of $m^2_G$.  Now, only
configurations with $e^2 \Phi({\mathcal A}) \in {\mathbb Z}$ may have a
finite action, and this implies that $\xi$ could have only been
${e^2}/{2\pi}$ or an integer multiple of this quantity, i.e.,
\begin{equation}\label{eq:quant}
m_G^2 \,=\, k \,\frac{e^2}{2\pi} \;,\;\;\; k\,=\, 0,1,2,\ldots 
\end{equation}
  
It must of course be possible to prove this quantization, starting
from the fermionic side of the problem.  Indeed, this is the case
for the one-loop approximation, since we also know that, to that
order, the integral of the anomaly is in fact the index of the Dirac
operator, namely:
\begin{equation}\label{eq:index}
\int d^2x \, {\mathcal G}({\mathcal A}) \;=\; \frac{1}{2 \pi}  (n_+ - n_-)
\end{equation}
where $n_\pm$ denotes the number of zero modes of positive and negative
chirality in the given background. In our conventions, it is easy to
realize that \mbox{$n_+ - n_- = e^2 \Phi({\mathcal A})$}.

In the case in which a current-current interaction is introduced, the corresponding effect can be evaluated by means of the bosonization rules, which give 
\begin{equation}\label{eq:cc}
\partial_\mu \langle J_\mu^5 \rangle \;=\; \frac{e^2}{\pi+g} \epsilon_{\mu\nu} \partial_{\mu} A_{\nu} \;\;\;;
\end{equation}
where $g$ denotes the current-current coupling constant. This relationship would imply, through equation ({\ref{eq:anom2}}), a renormalized gauge invariant mass, in agreement with the result of ref.\cite{glr}.

Also, when the fermions are massive, the right hand side of
equation ({\ref{eq:defanom}}) has to be supplemented with the
additional term $-2m {\bar\psi}\gamma_5\psi$ coming from the explicit chiral
symmetry breaking. Accordingly, equation ({\ref{eq:anom1}}) reads
\begin{equation}\label{eq:anom10}
\int d^2x \; \partial_\mu \langle J_\mu^5 \rangle\;=\;\int d^2x \;\left( {\mathcal G}({\mathcal A})
-2m \langle {\bar\psi}\gamma_5\psi\rangle \right)
 \;=\; \lim_{k\to0} {\tilde \Pi}(k^2) \;\;
\lim_{k\to0} \epsilon_{\mu\nu} i k_\mu {\tilde {\mathcal A}}_\nu (k) \;. 
\end{equation}
Due to the short ranged behavior of the massive Dirac fields, the left
hand side identically vanishes, which together with equation
(\ref{eq:pi0}) gives a zero gauge invariant mass $m_G$.  In
particular, this implies that the contribution of the chiral anomaly
cancels against that of the explicit chiral symmetry breaking
term~\cite{raj}.

\section{Momentum Space Representation for the Gauge Invariant Mass}
\begin{figure}[h]
\begin{center}
\begin{picture}(0,0)%
\includegraphics{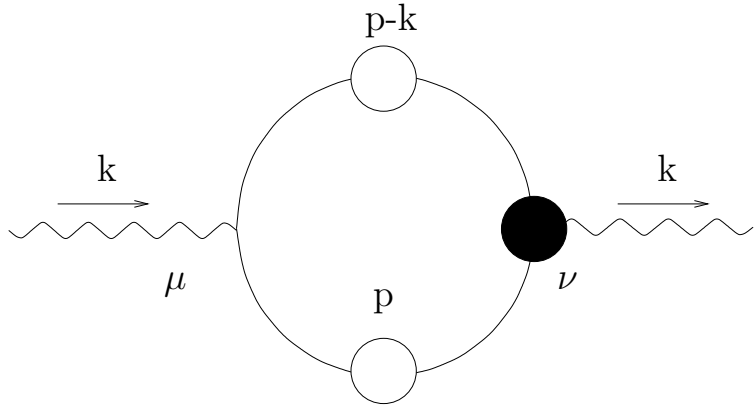}%
\end{picture}%
\setlength{\unitlength}{2763sp}%
\begingroup\makeatletter\ifx\SetFigFont\undefined%
\gdef\SetFigFont#1#2#3#4#5{%
  \reset@font\fontsize{#1}{#2pt}%
  \fontfamily{#3}\fontseries{#4}\fontshape{#5}%
  \selectfont}%
\fi\endgroup%
\begin{picture}(6703,3580)(1425,-4337)
\put(2251,-2311){\makebox(0,0)[lb]{\smash{\SetFigFont{14}{16.8}{\familydefault}{\mddefault}{\updefault}{\color[rgb]{0,0,0}k}%
}}}
\put(7276,-2311){\makebox(0,0)[lb]{\smash{\SetFigFont{14}{16.8}{\familydefault}{\mddefault}{\updefault}{\color[rgb]{0,0,0}k}%
}}}
\put(4651,-961){\makebox(0,0)[lb]{\smash{\SetFigFont{14}{16.8}{\familydefault}{\mddefault}{\updefault}{\color[rgb]{0,0,0}p-k}%
}}}
\put(6376,-3286){\makebox(0,0)[lb]{\smash{\SetFigFont{14}{16.8}{\familydefault}{\mddefault}{\updefault}{\color[rgb]{0,0,0}$\nu$}%
}}}
\put(2851,-3286){\makebox(0,0)[lb]{\smash{\SetFigFont{14}{16.8}{\familydefault}{\mddefault}{\updefault}{\color[rgb]{0,0,0}$\mu$}%
}}}
\put(4726,-3436){\makebox(0,0)[lb]{\smash{\SetFigFont{14}{16.8}{\familydefault}{\mddefault}{\updefault}{\color[rgb]{0,0,0}p}%
}}}
\end{picture}
\caption{The exact vacuum polarization graph.}
\label{fig:fig1}
\end{center}
\end{figure}

We try, in what follows, to use the fermionic point of view
exclusively, in an attempt to derive the previous results therefrom,
without assuming that a loop expansion has been performed.  To that
end, let us derive an exact expression for $m_G$ in terms of the 
fermion propagator.  We start from the ($D$-dimensional) general
expression for ${\tilde \Pi}_{\mu\nu}$, as given by the Schwinger-Dyson
equations~\cite{itz}, which amount to the diagram of
Figure~\ref{fig:fig1},
where the white blobs are included to mean that the lines are full
fermion propagators, while the black blob represents the full vertex
functions.  The external legs are of course to be truncated, but we
have drawn them for the sake of clarity. The analytic expression
corresponding to Figure~\ref{fig:fig1} is
\begin{equation}\label{eq:sde}
{\tilde \Pi}_{\mu\nu}(k) \;=\; - e^2 \, \int \frac{d^Dp}{(2\pi)^D}\;{\rm tr} 
\left[ \gamma_\mu \,S_F(p)\, \Gamma_\nu(p,p-k) \, S_F(p-k) \right]
\end{equation}
where $S_F$ is the momentum propagator and $\Gamma_\nu$ the exact vertex
function.  Noting that $\Pi_{\mu\nu}$ may be expressed as in
(\ref{eq:pimn}), we see that
\begin{equation}
m^2_G \,=\, {\tilde \Pi}(0) \,=\, (D-1)^{-1} \lim_{k\to0} \, {\tilde \Pi}_{\mu\mu}(k) \;, 
\end{equation}
or
\begin{equation}\label{eq:mg1}
m^2_G \;=\; - e^2  (D-1)^{-1} \lim_{k\to0}  
\int \frac{d^Dp}{(2\pi)^D}\, {\rm tr} \left[ \gamma_\mu \, S_F(p) \, \Gamma_\mu(p,p-k) \, 
S_F(p-k) \right]\;.
\end{equation}
Since the theory is gauge invariant, we may relate $\lim_{k\to0}
\Gamma_\mu(p,p-k)$ to the fermion propagator, by using the exact Ward
identity:
\begin{equation}\label{eq:wid}
\lim_{k\to0} \Gamma_\mu(p,p-k) \,=\, -i \frac{\partial}{\partial p_\mu} S_F^{-1} (p)\;.
\end{equation}
This identity may be used in (\ref{eq:mg1}) to obtain 
$$
m^2_G \;=\; i e^2 \, (D-1)^{-1} \int \frac{d^Dp}{(2\pi)^D}\, {\rm tr}
\left[ \gamma_\mu S_F(p) \frac{\partial}{\partial p_\mu} S_F^{-1} (p) S_F(p) \right]\;
$$
\begin{equation}\label{eq:mg2}
=\, - i e^2 \,(D-1)^{-1}\, \int \frac{d^Dp}{(2\pi)^D}\; \frac{\partial}{\partial p_\mu}\, {\rm tr} 
\left[ \gamma_\mu  S_F(p) \right]\;,
\end{equation}
which is an expression we shall consider in detail below.  It is still
formal, in the sense that we have not yet made explicit any
regularization method.

\begin{figure}[h]
\begin{center}
\begin{picture}(0,0)%
\includegraphics{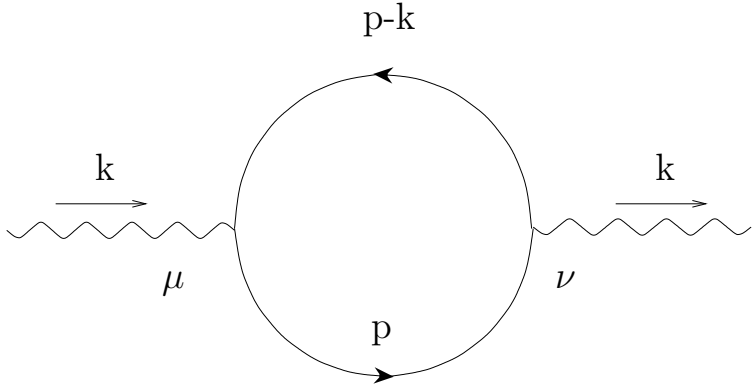}%
\end{picture}%
\setlength{\unitlength}{2763sp}%
\begingroup\makeatletter\ifx\SetFigFont\undefined%
\gdef\SetFigFont#1#2#3#4#5{%
  \reset@font\fontsize{#1}{#2pt}%
  \fontfamily{#3}\fontseries{#4}\fontshape{#5}%
  \selectfont}%
\fi\endgroup%
\begin{picture}(6703,3395)(1425,-4152)
\put(2251,-2311){\makebox(0,0)[lb]{\smash{\SetFigFont{14}{16.8}{\familydefault}{\mddefault}{\updefault}{\color[rgb]{0,0,0}k}%
}}}
\put(7276,-2311){\makebox(0,0)[lb]{\smash{\SetFigFont{14}{16.8}{\familydefault}{\mddefault}{\updefault}{\color[rgb]{0,0,0}k}%
}}}
\put(4726,-3736){\makebox(0,0)[lb]{\smash{\SetFigFont{14}{16.8}{\familydefault}{\mddefault}{\updefault}{\color[rgb]{0,0,0}p}%
}}}
\put(4651,-961){\makebox(0,0)[lb]{\smash{\SetFigFont{14}{16.8}{\familydefault}{\mddefault}{\updefault}{\color[rgb]{0,0,0}p-k}%
}}}
\put(6376,-3286){\makebox(0,0)[lb]{\smash{\SetFigFont{14}{16.8}{\familydefault}{\mddefault}{\updefault}{\color[rgb]{0,0,0}$\nu$}%
}}}
\put(2851,-3286){\makebox(0,0)[lb]{\smash{\SetFigFont{14}{16.8}{\familydefault}{\mddefault}{\updefault}{\color[rgb]{0,0,0}$\mu$}%
}}}
\end{picture}
\caption{The one-loop vacuum polarization graph.}
\label{fig:fig2}
\end{center}
\end{figure}

Before exploring its divergences in the general case, we note that, to
one loop order, they are the well-known divergences in the vacuum
polarization graph of Figure~\ref{fig:fig2}.

\begin{figure}[h]
\begin{center} 
\begin{picture}(0,0)%
\includegraphics{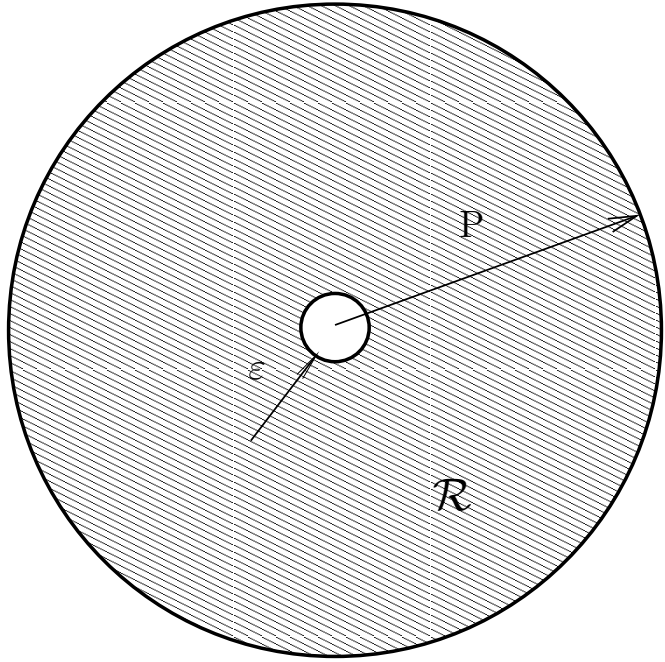}%
\end{picture}%
\setlength{\unitlength}{2901sp}%
\begingroup\makeatletter\ifx\SetFigFont\undefined%
\gdef\SetFigFont#1#2#3#4#5{%
  \reset@font\fontsize{#1}{#2pt}%
  \fontfamily{#3}\fontseries{#4}\fontshape{#5}%
  \selectfont}%
\fi\endgroup%
\begin{picture}(5626,5624)(3038,-5933)
\put(6706,-4651){\makebox(0,0)[lb]{\smash{\SetFigFont{17}{20.4}{\familydefault}{\mddefault}{\updefault}{\color[rgb]{0,0,0}${\mathcal R}$}%
}}}
\put(6931,-2311){\makebox(0,0)[lb]{\smash{\SetFigFont{14}{16.8}{\familydefault}{\mddefault}{\updefault}{\color[rgb]{0,0,0}P}%
}}}
\put(5131,-3526){\makebox(0,0)[lb]{\smash{\SetFigFont{14}{16.8}{\familydefault}{\mddefault}{\updefault}{\color[rgb]{0,0,0}$\varepsilon$}%
}}}
\end{picture}
\caption{The integration region for the integrals in (\ref{eq:mg5}).}
\label{fig:fig3}
\end{center}
\end{figure}

In (1+1) the expression for $m_G$ suffers from both UV and IR
divergences, as it is clear from (\ref{eq:mg2}), since, when applied
to the free Dirac propagator \mbox{$S_F^{(0)} = -i {\not \! p}^{-1}$},
it yields:
\begin{equation}\label{eq:mg4}
m^2_G\;=\; -  2 e^2 \int \frac{d^2p}{(2\pi)^2}\,\frac{\partial}{\partial p_\mu} (\frac{p_\mu}{p^2}) \;.
\end{equation}
In order to make sense of this expression, we exclude a circle of
radius $\varepsilon$ around the origin of the momentum plane to avoid the IR
singularity, and also use a Pauli-Villars regulator to tame the UV
divergences.  This amounts to defining
\begin{equation}\label{eq:mg5}
m^2_G\;=\; - 2 e^2 \int_{{\mathcal R}} \frac{d^2p}{(2\pi)^2}\,\frac{\partial}{\partial p_\mu} (\frac{p_\mu}{p^2}) 
+ 2 e^2  \int_{{\mathcal R}} \frac{d^2p}{(2\pi)^2}\,\frac{\partial}{\partial p_\mu} (\frac{p_\mu}{p^2+\Lambda^2})\;,
\end{equation}
where ${\mathcal R}$ is the region illustrated in
Figure~\ref{fig:fig3}, and $\Lambda$ denotes the mass of the regulator
field. The momentum $P$ denotes the radius of the integration region,
and of course $P \to \infty$ since the theory is already regularized.

By a straightforward application of the two dimensional Gauss' theorem, we
may convert these integrals of divergences into fluxes of radial
vector fields.  This procedure yields
\begin{equation}\label{eq:mg6}
m^2_G\;=\; - \frac{2 e^2}{(2\pi)^2} \,  \left[ - 2 \pi \frac{\varepsilon}{\varepsilon} + 2 \pi \frac{\varepsilon}{\varepsilon^2+\Lambda^2} \right]
\end{equation}
which in the $\varepsilon \to 0$ limit becomes:
\begin{equation}\label{eq:mg7}
m^2_G\;=\;  \frac{e^2}{\pi} \;, 
\end{equation}
as it should be.  The procedure we have followed has a simple
electrostatic analogy: $m_G$ is given by the integral of the
divergence of an `electric field', so by following that analogy it is
proportional to the total `electric charge'. Because of the regulator,
though, only the charges at the origin are relevant; namely, the
subtraction due to the UV regulator leads to:
\begin{equation}\label{eq:mg8}
m^2_G \;=\; \frac{i e^2}{(2 \pi)^2} \; \oint_{{\mathcal C}(\varepsilon)} \, dl \, {\hat n}_\mu 
{\rm tr}\left[\gamma_\mu S_F(p) \right]   
\end{equation}
where the integral is taken along a small curve of radius $\varepsilon$
enclosing the origin. ${\hat n_\mu}$ denotes the outer normal to
${\mathcal C}(\varepsilon)$. In what follows we shall argue that this kind of
expression can be generalized to the full theory. Indeed, we can
always say that the role of {\em any\/} UV regularization will be to
modify the large momentum behavior of the propagator, in such a way that 
all the points with infinite momentum  can be identified. For the 
Pauli-Villars case, this can be shown to hold simply by combining the 
contributions of both fermion propagators into one integral, to define a
regularized propagator. Namely,
\begin{equation}\label{eq:mg9}
m^2_G\;=\; - 2 e^2 \int_{{\mathcal R}} \frac{d^2p}{(2\pi)^2}\,
\frac{\partial}{\partial p_\mu} 
\left[\frac{\Lambda^2}{p^2 (p^2 + \Lambda^2)} p_\mu \right] \;,
\end{equation}
which contains the divergence (in momentum space) of a vector field
which decreases as $\sim p^{-4}$ at infinity.  

This will be one of the properties we shall demand of a regularization
at any finite order of the loop expansion; namely, the points at infinity in the momentum space
may, from the point of view of this calculation, be identified. The
other condition is that, in the small momentum region, i.e.\ momenta much
smaller than the cutoff, the behavior of the propagator should be the
same as for the unregularized propagator. Thus only the small region
around zero may contribute.  This, in turn, will produce a
non-vanishing answer only when the field is massless, as we shall see
now.

Assuming a standard Dirac action for the fermions, the general form of
the fermion propagator in momentum space is, of course,
\begin{equation}
S_F^{-1} (p) \;=\; i A(p) \not \! p + B(p) 
\end{equation}
where $A$ and $B$ are real functions that depend only on the scalar $p^2$. Then
an application of (\ref{eq:mg2}) yields:
$$
m^2_G \;=\; -i e^2 \, \int \frac{d^2p}{(2\pi)^2}\, \frac{\partial}{\partial p_\mu} {\rm tr}
\left[ \gamma_\mu \frac{A(p) \not \! p}{ A^2(p) p^2 + B^2(p) } \right]\;,
$$
\begin{equation}\label{eq:mg10}
\;=\; \frac{e^2}{(2 \pi)^2} \;{\rm tr}(I)\, \oint_{{\mathcal C}(\varepsilon)} \, dl \, {\hat n}_\mu 
\left[\frac{A(p) p_\mu}{A^2(p) p^2 + B^2}\right]   
\end{equation}
where ${\rm tr}(I)$ counts the dimension of the Dirac algebra
representation.  Evaluating the integral along ${\mathcal C}(\varepsilon)$,
\begin{equation}\label{eq:mg11}
m^2_G \;=\; \frac{e^2}{2 \pi} \;{\rm tr}(I)\;
\frac{A(\varepsilon) \varepsilon^2}{A^2(\varepsilon) \epsilon^2 + B^2(\varepsilon)}  \;.
\end{equation}
Since $\varepsilon \to 0$, we may conclude from here that, if the fermion is massive, 
$B(0) \neq 0$, then $m_G=0$. This is indeed the case for the
massive Schwinger model. Regarding the massless case, we obtain:
\begin{equation}\label{eq:mg12}
m^2_G \;=\; \frac{e^2}{2 \pi} \;{\rm tr}(I)\, \frac{1}{A(0)} \;=\;
\frac{e^2}{2 \pi} \;{\rm tr}(I)  \;,
\end{equation}
where we have used the normalization condition $A(0)=1$, which holds at any finite order of the loop expansion. This condition fixes the
residue of the perturbative electron propagator at the pole $p^2=0$. Strictly speaking, this normalization will change 
in an interacting theory. Indeed, the fermion propagator can in 
general, be rewritten by using the spectral representation~\cite{itz}:
\begin{equation}\label{eq:spect1}
S_F(p)\;=\; \int_0^\infty d\mu^2 \, \frac{- i  \rho_1(\mu^2) \not \! p \,+ \, \rho_2(\mu^2)}{p^2 \,+\, \mu^2}
\end{equation}
where $\rho_1$ and $\rho_2$ are real functions. It is then clear, by the linearity of 
this expression, that there will be a finite gauge invariant mass if there is  
an isolated pole at zero momentum, namely, if we can write:
\begin{equation}\label{eq:spect2}
S_F(p)\;=\; Z_2 \,\frac{- i \! p }{p^2}\,+\,
 \int_{m^2}^\infty d\mu^2 \, \frac{- i  \rho_1(\mu^2) \not \! p \,+ \, \rho_2(\mu^2)}{p^2 \,+\, \mu^2}
\end{equation}
where $m^2$ is the multiparticle threshold. In this situation, we would obtain,
\begin{equation}\label{eq:mg13}
m^2_G \;=\; \frac{e^2}{2 \pi} \;{\rm tr}(I)\, \frac{1}{A(0)} \;=\;
\frac{e^2}{2 \pi} \;{\rm tr}(I) Z_2 \;,
\end{equation}
where $Z_2$ is a constant, smaller that $1$ because of the spectral condition:
\begin{equation}\label{eq:spect3}
1 \;=\; Z_2 \,+\, \int_{m^2}^\infty d\mu^2 \rho_1(\mu^2) \;.
\end{equation}
Equation (\ref{eq:mg13}) may seem to contradict our remarks on the relation between 
$m_G^2$ and the anomaly, for example (\ref{eq:quant}). The resolution of this apparent 
paradox is that the anomaly, as we understood it in (\ref{eq:defanom}), is defined in 
terms of the divergence of the current. The matrix elements of this current, when 
evaluated through the reduction formulae, will require the introduction of a
$Z_2$ factor ($Z_2^{1/2}$ for each fermionic field), thus the proper relation that 
generalizes (\ref{eq:quant}) is:
\begin{equation}\label{eq:quant1}
m_G^2 \;=\; k \, \frac{e^2}{2\pi} \, Z_2 \;,\;\;\; k = 0,1,2, \ldots
\end{equation}

The original expression for $m_G$ required a
regularization procedure in order to be well defined. However, being
determined by the low momentum behavior of the propagator, $m_G$
should be independent of the regularization. For example, for a
generalized Pauli-Villars regularization in the one-loop case, one
might include a set of $N$ regulator fields, and define:
\begin{equation}\label{eq:regul1}
S_F^{reg}(p)\;=\; \sum_{n=0}^N \, C_n \frac{1}{i \not \! p + M_n} 
\end{equation}
where $M_n$ denote the regularizing masses, and the index $n=0$ is
reserved for the unregularized propagator, which has $M_0=0$ and
$C_0=1$. The masses $M_n$ and the coefficients $C_n$ for $n>0$ are
chosen in order to verify the desired behavior in the UV\@. However,
since only the behavior around zero momentum is relevant, and all the
regulators are massive, then just the unregularized massless
propagator contributes.

\section{Applications to Different Geometries  and Space-Time Dimensions}

\subsection{The case of the $1+1$ dimensional square lattice}  
It should be clear from what we have said above that the geometry of
the momentum space is crucial in the problem of evaluating $m_G$.  An
interesting example of this is the case of a lattice regularized
theory.  Indeed, assuming that the coordinate space has been
discretized to an (infinite) square lattice with lattice spacing $a$,
the momentum space becomes a (continuous) torus. The lattice points
are defined then as the set of points $\sigma_\mu = a \, t_\mu$ with $t_\mu \in
{\mathbb Z}$ for $\mu=0,1$.

The naive (i.e., no Wilson term) lattice fermion propagator for a massive
Dirac fermion then becomes: 
\begin{equation}\label{eq:fplat}
 S_F (p) \;=\; \frac{a}{i \gamma_\mu C_\mu(a p) + a m} 
\end{equation}
where $C_\mu(a p) = \sin(a p_\mu)$, and the $p_\mu$ are continuous variables (because the
lattice is infinite), in the first Brillouin zone, namely:
\begin{equation}
- \pi \,<\, p_\mu \,\leq\, \pi \;\;,\;\;\; \mu \,=\, 0,\,1\;.
\end{equation}
Then, the expression for $m_G$ is:
\begin{equation}
m^2_G \;=\; -i e^2 \, \int_{\mathcal B}  \frac{d^2p}{(2\pi)^2}\, \frac{\partial}{\partial p_\mu} {\rm tr}
\left[ \gamma_\mu  \frac{a}{i \not \!\! C(a p) + a m} \right]\;,
\end{equation}
where ${\mathcal B}$ denotes the first Brillouin zone. The fact that
we are dealing with a momentum torus is obvious because of the
periodicity of the functions $C_\mu$, a property which is preserved of
course for more general propagators. But the functions $C_\mu$ will
simultaneously vanish for the $(p_0,p_1)$ values in the set
\begin{equation}
\left\{ (0,0) \,,\,(0,\pi) \,,\, (\pi,0)\,,\, (\pi,\pi) \right\}\;,
\end{equation}
which contain not just the origin but also three other unwanted points.
The behavior of the lattice propagator is identical for each of these points, 
and equal to the one in the continuum propagator.  Indeed,
$C_\mu(ap) \sim a (p-q^\alpha)_\mu$ where $q^\alpha$ is any of the points 
where both $C_\mu$'s vanish. 

Then, when $m=0$, an application of the Gauss law on the torus, with
the circles around the four poles of the propagator excluded yields
four times the contribution of the physical pole at zero. This is
because of the contribution of the `charges' at the doublers. This is
just a manifestation of the `doubling' problem of lattice fermions,
related to the Nielsen-Ninomiya theorem~\cite{kn:Karsten.}.

Thus,
\begin{equation}
[m^2_G]_{lattice} \;=\; 4 \frac{e^2}{\pi} \;.
\end{equation}

\subsection{$QED_3$ at Finite Temperature}
Another circumstance where the structure of momentum space allows for
the emergence of a non-trivial gauge invariant mass, is the case of
$QED$ in $2+1$ dimensions at finite temperature in the presence of
`large' $A_0$ field configurations~\cite{ait}.  Parity conserving
versions of $QED(2+1)$, like the model introduced by Dorey and
Mavromatos~\cite{dm} have been extensively studied as quantum field
theory models at finite temperature~\cite{ait}. Most of what we shall
say about this here, however, holds true both for the parity
conserving and the parity breaking cases.

We now discuss $QED(2+1)$ with regards only to one of its aspects,
namely, its gauge invariant mass.  We recall that in finite
temperature quantum field theory Lorentz invariance is lost.  Indeed,
in the Matsubara formalism, which we shall adopt, the time coordinate
runs from $0$ to $\beta=\frac{1}{T}$.  This lack of Lorentz invariance
implies that one shall, in principle, have different masses for the
spatial and temporal components of the gauge field. Namely, the
natural extension of (\ref{eq:defmt}) to this case would be
\begin{equation}\label{eq:defmtn0}
{\mathcal L}_m \;=\; \frac{1}{2} m^2_{el} \, A_0(x,\tau) A_0(x,\tau) 
\;+\; \frac{1}{2} m^2_{mag} \, A_j(x,\tau) \delta^\perp_{jk} A_k(x,\tau)  
\end{equation}
where 
$\delta^\perp_{jk}=\delta_{jk}-\frac{\partial_j\partial_k}{{\mathbf \partial}^2}$ 
is the transverse Kronecker $\delta$ for the $2$ spatial dimensions, and
$\tau$ is the imaginary time. 
The two components $m_{el}$ and $m_{mag}$ are the electric and magnetic masses, 
which can of course be different for $T\neq0$. These two masses play the role
of components of a `mass tensor' for $A_\mu$.  
We shall deal exclusively with the magnetic mass. It is not difficult to apply 
a similar derivation to the one used for the $T=0$ case, to obtain
\begin{equation}
m^2_{el}\;=\; -i e^2 \frac{1}{\beta} \, \sum_{n=-\infty}^{+\infty} 
\int \frac{d^2p}{(2 \pi)^2} \, \frac{\partial}{\partial p_j} 
{\rm tr} \left[\gamma_j S_F(p,n) \right]
\end{equation}
where the sum runs over the Matsubara frequencies $\omega_n = (2 n + 1) \pi T$,
and $S_F$ is the finite temperature fermion propagator. 

For the free, massless fermion propagator in the presence of a large $A_0$,
we have~\cite{ait}:
\begin{equation}
S_F(p,n)\;=\; \frac{1}{i \gamma_0 {\tilde \omega}_n + i \gamma_k p_k}
\end{equation}
where ${\tilde\omega_n} = \omega_n + e A_0$. $A_0$ is assumed to be constant, a
fact what can always be achieved by a small gauge transformation. The
value of this constant is of course determined by the quantity $\int d\tau
A_0(\tau)$, where $A_0(\tau)$ is an arbitrary time dependent configuration.
Then,
\begin{equation}
m^2_{el}\;=\; -i e^2 \frac{1}{\beta} \, {\rm tr}(I) \, 
\sum_{n=-\infty}^{+\infty} \int \frac{d^2p}{(2 \pi)^2} \, 
\frac{\partial}{\partial p_j} \frac{p_j}{{\mathbf p}^2 + {\tilde \omega_n}^2} 
\end{equation}
which looks like a series of $1+1$ dimensional contributions. Of
course, whenever ${\omega}_n = - e A_0$, we have a non-vanishing
contribution, of the same kind as those appearing in $1+1$ dimensions.
That condition on $A_0$ amounts to
\begin{equation}
e \int_0^\beta d\tau A_0(\tau) \;=\; (2 n + 1) \pi \;. 
\end{equation}  
Thus, we may write the result for $m^2_{el}$ as:
\begin{equation}
m^2_{el}\;=\;  \frac{e^2}{2 \pi \beta} \, {\rm tr}(I) \,
\sum_{l=-\infty}^{+\infty}  
\delta [e \int_0^\beta d\tau A_0(\tau) -  (2 n + 1) \pi ] \;.
\end{equation}

The $\beta$ dependence should have been expected by dimensional analysis,
since in $2+1$ dimensions $e^2$ has the units of a mass. Regarding the
existence of particular points where the magnetic mass is generated,
the physical reason for that is that when the condition for $A_0$ is
met, dimensional reduction occurs because there is a massless mode,
and the reduced model is then tantamount to a Schwinger model.

\section{Conclusions}\label{sec:conc}

In this work a useful momentum space representation for the gauge
invariant mass has been obtained and applied to different situations,
namely: massless and massive $QED$ in $1+1$ dimensions, the $1+1$
dimensional square lattice and $QED_3$ at finite temperature. The key
ingredient for this representation is the validity of gauge invariance
expressed by the Ward identity, and of course its consistency with the
dynamics defined by the Schwinger-Dyson equations.

It should be clear from (\ref{eq:mg2}) that the value of $m^2_G$
depends entirely on the infrared behavior of the fermion propagator,
resulting indeed from the infrared dynamics.  In particular, the
momentum representation for the gauge invariant mass in $1+1$ dimensions
turns out to be related to the chiral anomaly. This relationship
suggests that $m_G$ should possess some kind of stability against
perturbations. For instance, the introduction of a current-current
interaction amounts to a smooth change~\cite{glr} of the coefficient
$A(0)$ in equation (\ref{eq:mg12}), while preserving the vanishing of
$B(p)=0$. This means that, apart from a normalization factor, $m^2_G$
is still finite and non-vanishing.

A further comment we would like to add, and which is related to the
previous discussion is one concerning the meaning of formula
(\ref{eq:mg2}). The main difference between our momentum space
representation for $m_G$ and similar representations~\cite{mats,mats1,volovik} 
lies in the presence of the
factor $A(0)$. This is related to the fact that in our case the
formula cannot be written just in terms of the SU(2) projection of the
fermion propagator as in the other cases. Therefore, when including
interactions $m_G$ is not protected from being renormalized. In this
sense, the quantized gauge invariant mass of the $1+1$ dimensional
Schwinger model is not stable against interactions. In the present case, what is stable against perturbations is the
ratio between $m_G$ and $Z_2$. If perturbations are included in the
massive Schwinger model or in the massless case, $m_G$ will however
still be $m_G=0$ or $m_G\neq 0$, respectively.

We finish with a comment and outlook on the possible implications of
our results to higher dimensional systems at zero temperature, where
dimensional reduction does not occur. It should be noted that the main
difficulty is the fact that $m_G^2$, as given by (\ref{eq:mg2}), seems
to be not interesting for the higher dimensional case, because of the
IR behavior of the fermion propagators. One possible way out of this,
could be to consider non-standard fields like, for instance, dipole fields. 
This, however, should require a re-derivation of the main
results, since the structure of the model will, in principle, be
different from the standard minimal coupling case. Results on this
will be reported elsewhere.

\section*{Acknowledgements}

The Conselho Nacional de Desenvolvimento Cient{\'\i}fico e
Tecnol{\'o}gico CNPq-Brazil, the Funda{\c{c}}{\~{a}}o de Amparo
{\`{a}} Pesquisa do Estado do Rio de Janeiro (Faperj), Fundaci{\'o}n
Antorchas (Argentina), and the SR2-UERJ are acknowledged for their
financial support.


\end{document}